\documentclass[final,5p,times,twocolumn]{elsarticle}
\pdfoutput=1

 \usepackage{hyperref}

 \usepackage{graphicx}

\usepackage{amssymb}

\graphicspath{{img/}}

\journal{Vacuum}

\begin{document}

\begin{frontmatter}



\title{Improved model for transmission probabilities of membrane bellows based on TPMC simulations}


\author[adrlabel3,adrlabel4]{M. Krause}
\author[adrlabel4]{J. Wolf\corref{cor1}}
\cortext[cor1]{corresponding author: marcel.krause@kit.edu and joachim.wolf@kit.edu}

\address[adrlabel3]{Institute for Theoretical Physics (ITP), Karlsruhe Institute of Technology (KIT), Wolfgang-Gaede-Str. 1, 76131 Karlsruhe, Germany}
\address[adrlabel4]{Institute of Experimental Particle Physics (ETP), Karlsruhe Institute of Technology (KIT), Wolfgang-Gaede-Str. 1, 76131 Karlsruhe, Germany}

\begin{abstract}
Many complex vacuum systems include edge-welded bellows. Their simulation in the molecular flow regime with a Test Particle Monte Carlo (TPMC) code, such as {\texttt{MolFlow+}}, can take considerable amounts of computing power and time. Therefore we investigated the change of the transmission probability of a bellow compared to a cylindrical tube in TPMC simulations. The results were used to develop an empirical model to simplify the geometry of a bellow in a TPMC model by replacing it with a cylindrical tube of extended length, consequently compensating the reduced conductance of the bellow. The simulated transmission probabilities of a variety of different bellow lengths have been used to fit the parameters of the model. Each bellow geometry simulated with {\texttt{MolFlow+}} comprised two cylindrical tubes, one at each end, and the central bellow section. The geometry of each bellow is described by two quantities. The first one is the total length of the geometry model normalized to the inner diameter of the bellow. The second one is the fraction of the bellow section with regard to the total length. Simulating a tube instead of a long bellow reduced the simulation time by a factor of up to 1000, while the error introduced through the replacement of the bellow with a cylindrical tube of modified length was in most cases negligible.
\end{abstract}

\begin{keyword}
Edge-welded bellow \sep transmission probability \sep improved model \sep TPMC simulation 



\end{keyword}

\end{frontmatter}


\section{Introduction}\label{sec1-intro}

Edge-welded bellows are important and common components of vacuum systems. Due to multiple scattering of gas particles inside the bellow elements, the transmission probability of gas particles through a bellow can be significantly reduced in comparison to a cylindrical tube of the same length. Therefore, when aiming for a realistic simulation of a vacuum system, edge-welded bellows can not be neglected without a loss of accuracy of the simulation. On the other hand, the inclusion of edge-welded bellows significantly complicates the simulation. In a typical Test Particle Monte Carlo (TPMC) vacuum simulation, the geometry of all components of the vacuum system is approximated e.g. by a triangle mesh. When comparing a cylindrical tube with a bellow of the same length, the latter increases the amount of triangles needed easily by a few orders of magnitude, therefore increasing the computation time of the simulation considerably. The increased number of hits of the walls inside the narrow bellow elements and the resulting longer trajectories until the particle leaves the bellow, add to the computation time, too. To our knowledge, empirical formulae for the transmission probabilities of edge-welded bellows have not been presented in the literature so far, while for the case of cylindrical tubes, empirical formulae for arbitrary tube lengths exist. The aim of this paper is to provide an empirical formula for the transmission probability of edge-welded bellows of arbitrary length. By using this formula, we propose a method for replacing edge-welded bellows with cylindrical tubes of appropriately adjusted length in vacuum simulations, which leads to a reduction of simulation time without a significant loss of accuracy.

The paper is organized as follows. In Sec. \ref{sec:dynamicsRarefiedGases}, we shortly explain the dynamics of rarefied gases, present analytic formulae for transmission probabilities for cylindrical tubes as given in literature and extend the discussion to the case of edge-welded bellows. In Sec. \ref{sec:TPMCSimulationsBellows}, we give a short introduction to the TPMC software {\texttt{MolFlow+}} which is used for our TPMC simulations. We proceed to describe the principal method and relevant parameters of our bellow simulations. In Sec. \ref{sec:simulationResults}, we present the results of our simulations and discuss the dependence of the transmission probability on the bellow element angle. Additionally, we propose a new empirical model for the transmission probability of edge-welded bellows in dependence of their geometry parameters and analyze the increase in computation time when including edge-welded bellows in a vacuum simulation in comparison to a cylindrical tube. We complete the analysis with a proposal of a method for removing edge-welded bellows from vacuum simulations, therefore decreasing the time needed for a simulation, without losing the desired accuracy of the simulation.

\section{Dynamics of rarefied gases}\label{sec:dynamicsRarefiedGases}

\subsection{Flow of rarefied gases through cylindrical tubes}\label{sec:flowThroughTubes}

The physical problem of the flow of rarefied gases through cylindrical tubes has been analyzed both theoretically and experimentally for a long time, e.g. by Knudsen who proposed analytic formulae for the conductance of cylindrical tubes and the gas rate that flows through them\cite{lit:knudsen}. These formulae were corrected and improved by Smoluchowski\cite{lit:smoluchowski} and Clausing\cite{lit:clausing}. The latter additionally derived an analytic formula of the transmission probability $W$ of cylindrical tubes with arbitrary length, i.e. the probability of a gas particle to flow through the tube, in the form of an integral equation. For tubes with arbitrary length $l$ and diameter $d$, this integral can not be solved analytically in closed form, but it can be solved through numerical integration. For special configurations however, the integral is approximately solvable. For tubes that are very short compared to their diameter, Clausing found the analytic solution

\begin{equation}
	W \approx 1 - \frac{l}{d} \quad \quad (l \ll d) ~,
\label{eq:clausinFormula1}
\end{equation}

\noindent while for very long tubes with respect to their diameter, the well-known limit

\begin{equation}
	W \approx \frac{4}{3} \frac{d}{l} \quad \quad (l \gg d) 
\label{eq:clausinFormula2}
\end{equation}

\noindent applies. For tubes with lengths between these two limits, the empirical formula 

\begin{equation}
	W = \frac{14 + 4 \frac{l}{d}}{14 + 18 \frac{l}{d} + 3 \left( \frac{l}{d} \right) ^2}
\label{eq:joustenFormula}
\end{equation}

\noindent was derived \cite{lit:jousten}, which for arbitrary values of $l/d$ gives a good approximation to the analytic result up to a relative error below $0.6\%$. It is easy to verify that Eq. (\ref{eq:joustenFormula}) converges to Eqs. (\ref{eq:clausinFormula1}) and (\ref{eq:clausinFormula2}) for the respective limits of $l/d$.

\subsection{Flow of rarefied gases through edge-welded bellows}\label{sec:flowThroughBellows}

\begin{figure}[t]
\centering
\includegraphics[width=0.48\textwidth]{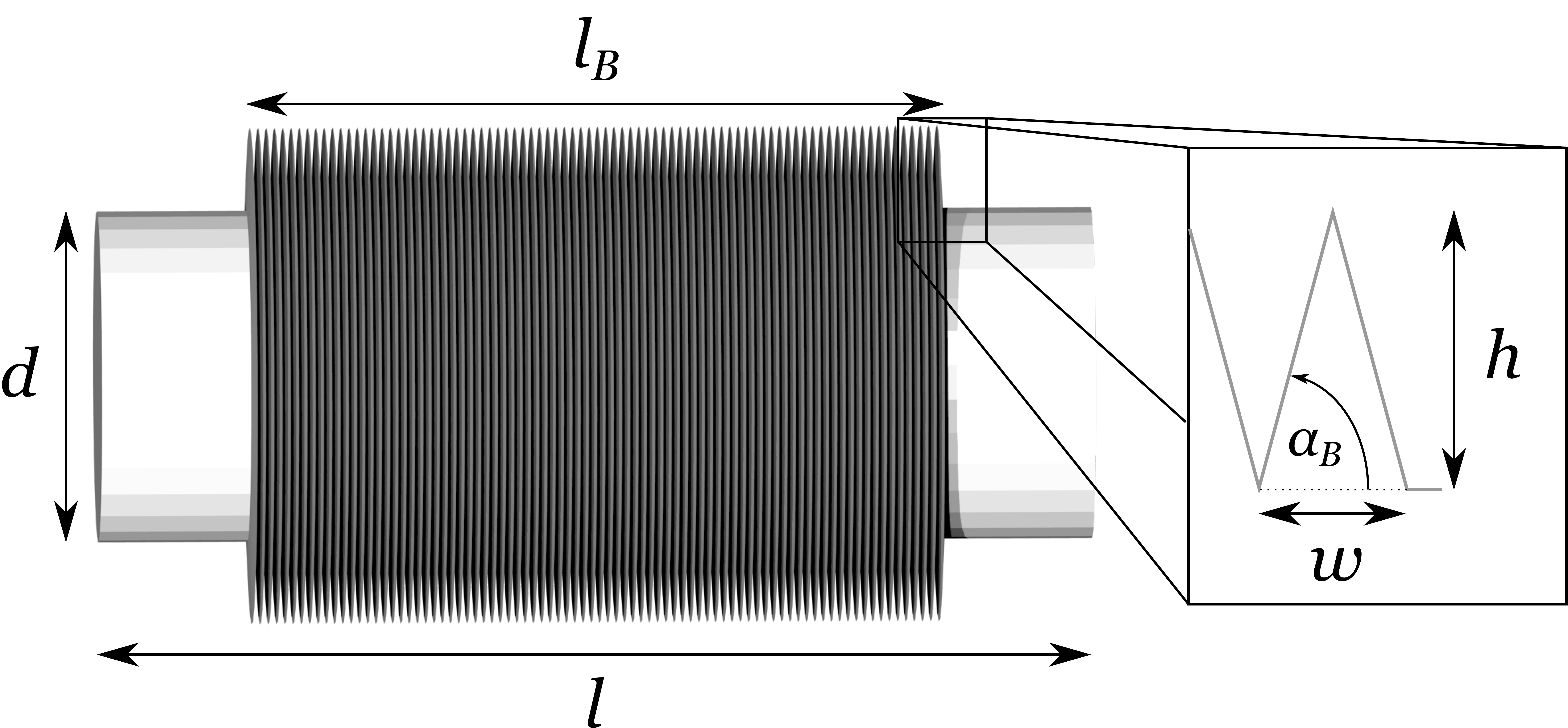} 
\caption{CAD model of an edge-welded bellow. The edge-welded bellow is parametrized by the length $l$ and diameter $d$ of the tube, the length of the bellow segment $l_B$, and the height $h$ and width $w$ of a bellow element. The bellow angle $\alpha_B$ is determined via the ratio $h/w$.} \label{fig:bellowSchematic} 
\end{figure}

The geometry of edge-welded bellows is significantly more complex than that of cylindrical tubes. Depicted in Fig. \ref{fig:bellowSchematic} is a computer-aided design (CAD) model of a typical edge-welded bellow, consisting of two cylindrical tubes on both ends and the actual bellow element in between the two tubes\footnote{The geometry of hydroformed bellows, often used for systems where less flexibility is required, differs slightly from the triangular edge-welded CAD model presented in Fig. \ref{fig:bellowSchematic}. In first approximation, the model is still valid, since the exact shape of the bellow element has only a minor influence on the transmission probability.}. The geometry of the bellow is characterized by the total length of the tube $l$, the inner diameter of the tube $d$, the length $l_B$ of the bellow element and the height $h$ and width $w$ of a bellow segment, as depicted in the figure. Via the trigonometric relation 

	\begin{equation}
		\tan (\alpha _B ) = \frac{2h}{w} ~,
	\label{eq:bellowAngle}
	\end{equation}

\noindent the height of the bellow is directly connected to the opening angle $\alpha _B$ of the bellow segment. Gas particles that enter the bellow typically scatter numerous times inside the bellow segments. Depending on the geometry parameters of the bellow, this leads to an increased beaming effect towards the walls of the bellow or towards the inlet of the tube. As a consequence, edge-welded bellows have a lower transmission probability for gas particles than cylindrical tubes of the same length and dia\-meter. Therefore, bellows are important and non-negligible components of vacuum systems, especially when aiming for an accurate simulation of the system.

The integral formula for the transmission probability $W$ as derived by Clausing was quite complicated even for the simple case of cylindrical tubes with arbitrary length. Due to the more complex geometry of edge-welded bellows, it is therefore not expected that an analytic formula for the transmission probability can be derived from first principles, and neither is it expected that such a formula would have a simple analytic solution. In order to quantify the transmission probability for a bellow, we use an \textit{ansatz} that is motivated by the empirical form of Eqs. (\ref{eq:clausinFormula1}) and (\ref{eq:joustenFormula}):

\begin{equation}
	W = \frac{1 + c_1 \frac{l}{d}}{1 + c_2 \frac{l}{d} + c_3 \left( \frac{l}{d} \right) ^2} ~.
\label{eq:ourBellowFormula}
\end{equation}

\noindent The coefficients $c_i$ are \textit{a priori} undetermined and could be fixed by e.g. measurements or TPMC simulations. We want to point out that these coefficients depend on the other geometry parameters of the bellow. In the case of bellows used in actual vacuum systems, the bellow height and width are usually fixed. Hence, we choose the coefficients $c_i$ to depend on the length $l_B$ of the actual bellow element, only. As our analysis described in Sec. \ref{sec:dependenceElementGeometry} shows, this is a valid approximation for typical bellows. In order to determine the parameters $c_i$, we simulated several variations of bellows geometries as described in Sec. \ref{sec:methodBellowSimulations}.

\section{TPMC simulations of edge-welded bellows \label{sec:TPMCSimulationsBellows}}

Before we describe the method and parameters of our bellow simulations, we first explain the basic concepts of the TPMC program {\texttt{MolFlow+}}. The publicly available {\texttt{C++}} code has been developed and maintained by R. Kersevan et al. \cite{lit:kersevan} and is hosted by CERN \cite{lit:molflow}.

\subsection{Introduction to MolFlow+ \label{sec:MF_basics}}

The {\texttt{C++}} code {\texttt{MolFlow+}} is a TPMC program for the simulation of vacuum systems in the regime of molecular flow. The program allows both for the import of CAD geometries as well as the creation of models of vacuum systems inside {\texttt{MolFlow+}} itself. In the following, we only want to present the basic concepts of the program. A detailed description of the functionality of the software can be found in \cite{lit:MolFlowGuide}.

When importing a model of a vacuum system into {\texttt{MolFlow+}}, the program approximates the geometry of the model by dividing it into a polygonal mesh, with the individual elements of the mesh being called \textit{facets}. For each facet, the user can define its individual properties, e.g. the sticking coefficient $\alpha$ of the facet, its outgassing rate or the reflection behavior of the facet with respect to colliding gas particles. In the TPMC simulation, the trajectories of gas particles start from the facets that are chosen as gas sources, with a rate corresponding to the outgassing rate, and their path through the vacuum set-up is tracked via a ray-tracing algorithm. Collisions of rays with a facet correspond to the gas particle hitting the respective wall of the vacuum set-up. The simulation ends when the gas particle is absorbed with a probability of $\alpha$ by the corresponding facet. Otherwise, it is emitted again as another ray, usually following a cosine law. Three counters are assigned to each facet, counting the desorption, hit, and absorption of particles. The two main quantities, generated by the TPMC simulation, that are necessary for our analysis are:

\begin{figure}[t]
\centering
\includegraphics[trim=16mm 8mm 32mm 20mm,clip,width=0.48\textwidth]{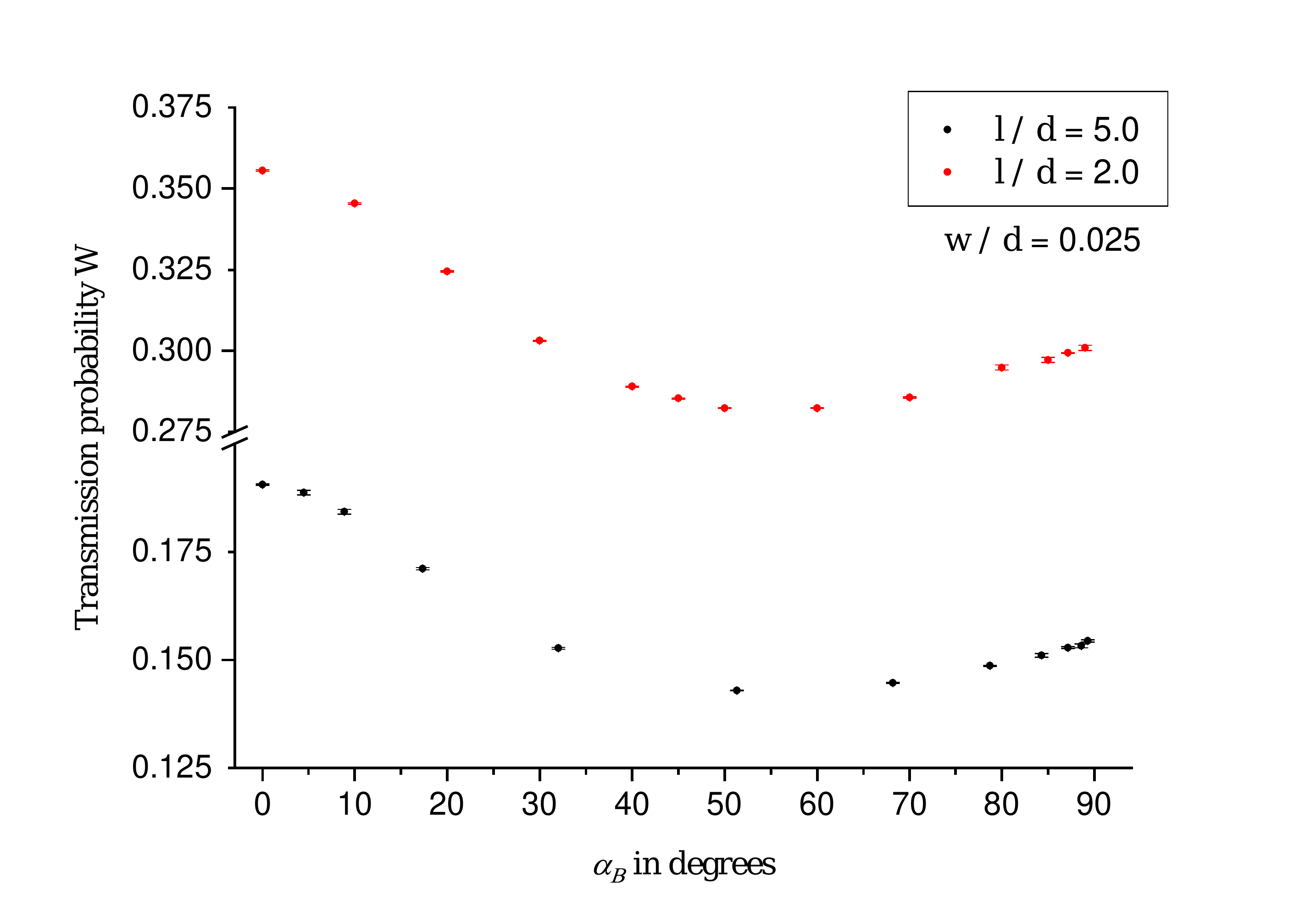} 
\caption{Results of the study on angle dependence. Shown is the transmission probability $W$ in dependence on the bellow angle $\alpha _B$ as defined in Eq. (\ref{eq:bellowAngle}) for two distinct, fixed values of $l/d$. For very small and large $\alpha _B$, $W$ is approximately independent of the bellow angle. The transmission probability for all $\alpha _B \gtrapprox 10^\circ$ is signficantly smaller than the one for $\alpha _B = 0$. For values near $\alpha _B \approx 55^\circ$, the transmission probability reaches a minimum.} \label{fig:resultsAngleDependence} 
\end{figure}

\begin{itemize}
\item \textbf{Desorptions $N_D$:} Total number of gas particles that are emitted from a source facet. In our simulation, particles are only desorbed from the inlet of the bellow.
\item \textbf{Absorptions $N_A$:} Total number of gas particles that are absorbed by a specific facet. In our simulations, the facet corresponding to the outlet of the bellow has a sticking coefficient $\alpha = 1$, equivalent to an ideal absorber, while the rest of the facets have $\alpha = 0$.
\end{itemize}

\begin{figure}[t] 
\centering
\includegraphics[width = 0.48\textwidth]{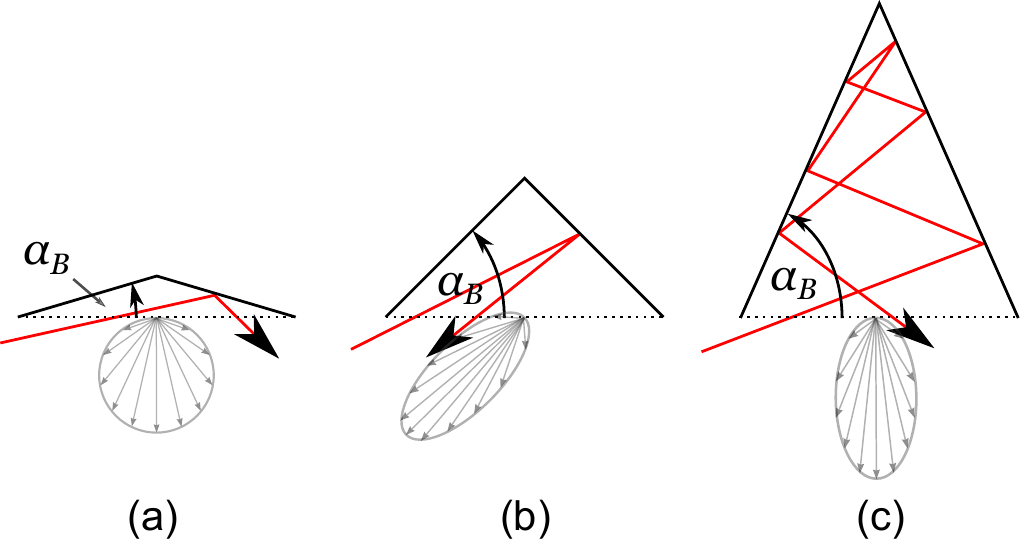} 
\caption{Sketch of a bellow element for (a) very small bellow angles $\alpha _B$, (b) angles $\alpha _B \approx 55^\circ$ where $W$ becomes minimal and (c) very large $\alpha _B$. Gas particles, whose paths are depicted by arrows in red, enter the bellow from the left and scatter on the walls of the bellow with distribution functions of the diffuse reflection as depicted by the gray ellipses and arrows. In case (a), the bellow is effectively a cylindrical tube and gas particles reflect by the cosine law, while in case (c), multiple scattering inside each bellow segment effectively makes it a gas source with increased beaming towards the inside walls of the bellow. In case (b), the backwards scattering for incoming particles on the bellow elements is maximal, leading to a backwards beaming effect and a minimal transmission probability of the bellow.} \label{fig:dependenceOnAngle} 
\end{figure}

\subsection{Method of the bellow simulations \label{sec:methodBellowSimulations}}

For the TPMC simulation of the transmission probabilities with {\texttt{MolfFlow+}} we constructed models of edge-welded bellows with a large variety of geometry parameters, as exemplarily depicted in Fig.~\ref{fig:bellowSchematic}. The models were designed using the open-source software {\texttt{Blender}}\cite{lit:blender}. The circular entrance facet was defined both as a gas source and as an ideal absorber, i.e. $\alpha = 1$. The outlet facet is only an ideal absorber. For all other facets, the sticking probability has been set to zero. This configuration corresponds to a set-up where gas particles enter the bellow only from one side, which is a valid approximation since we are only interested in the transmission probability of the bellow itself. The simulation of each bellow continued until the number of absorptions $N_A$ by the outlet facet was of the order $\mathcal{O} (10^5 - 10^6)$, ensuring sufficient statistical accuracy. This quantity, together with the total amount of emitted gas particles $N_D$, allows the calculation of the transmission probability $W$ by the formula

\begin{equation}
	W = \frac{N_A}{N_D} ~.
\end{equation}

\noindent We performed two studies on the transmission probabilities of the bellows, as described in the following:

\begin{enumerate}
	\item Qualitative analysis of the transmission probability $W$ in dependence on the geometry parameters of the single bellow segments. For this study, we only considered the case of a pure bellow, i.e. we set $l_B/l = 1.0$, and a bellow segment with a fixed width $w/d = 0.025$. For the design parameter $l / d$, two distinct, fixed values with $l/d \in \{ 2.0 ,\, 5.0 \}$ were simulated. For these two values, the height $h/d$ of the bellow element was varied between the values $h/d \in \left[ 0.000 ,\, 0.7161 \right]$ in 13 steps, corresponding to opening angles $\alpha _B \in \left[ 0^\circ ,\, 89^\circ \right]$. The results of this study are presented in Sec.~\ref{sec:dependenceElementGeometry}.

\item Quantitative analysis of the transmission probability $W$	in dependence of the total length $l$ and the length of the  bellow $l_B$. For this study, the geometry parameters of the bellow segments were fixed to the values $w/d = 0.025$ and $h/d = 0.25$, corresponding to an angle $\alpha _B \approx 87.14^\circ $. Both the total length of the tube $l/d \in \left[ 0.5 ,\, 20.0 \right]$ as well as the length of the bellow element compared to the total length of the tube $l_B / l \in \left[ 0.0 ,\, 1.0 \right]$ have been varied. For cylindrical tubes ($l_B/l = 1$) three additional lengths were simulated at $l/d$\, =\,50, 500, and 5000, increasing the reliability of the fitted parameters to very long tubes. The results of this study are presented in Sec.~\ref{sec:NewModelBellows}.
\end{enumerate}

\begin{figure}[h]
\centering
\includegraphics[trim=24mm 5mm 31mm 20mm,clip,width=0.48\textwidth]{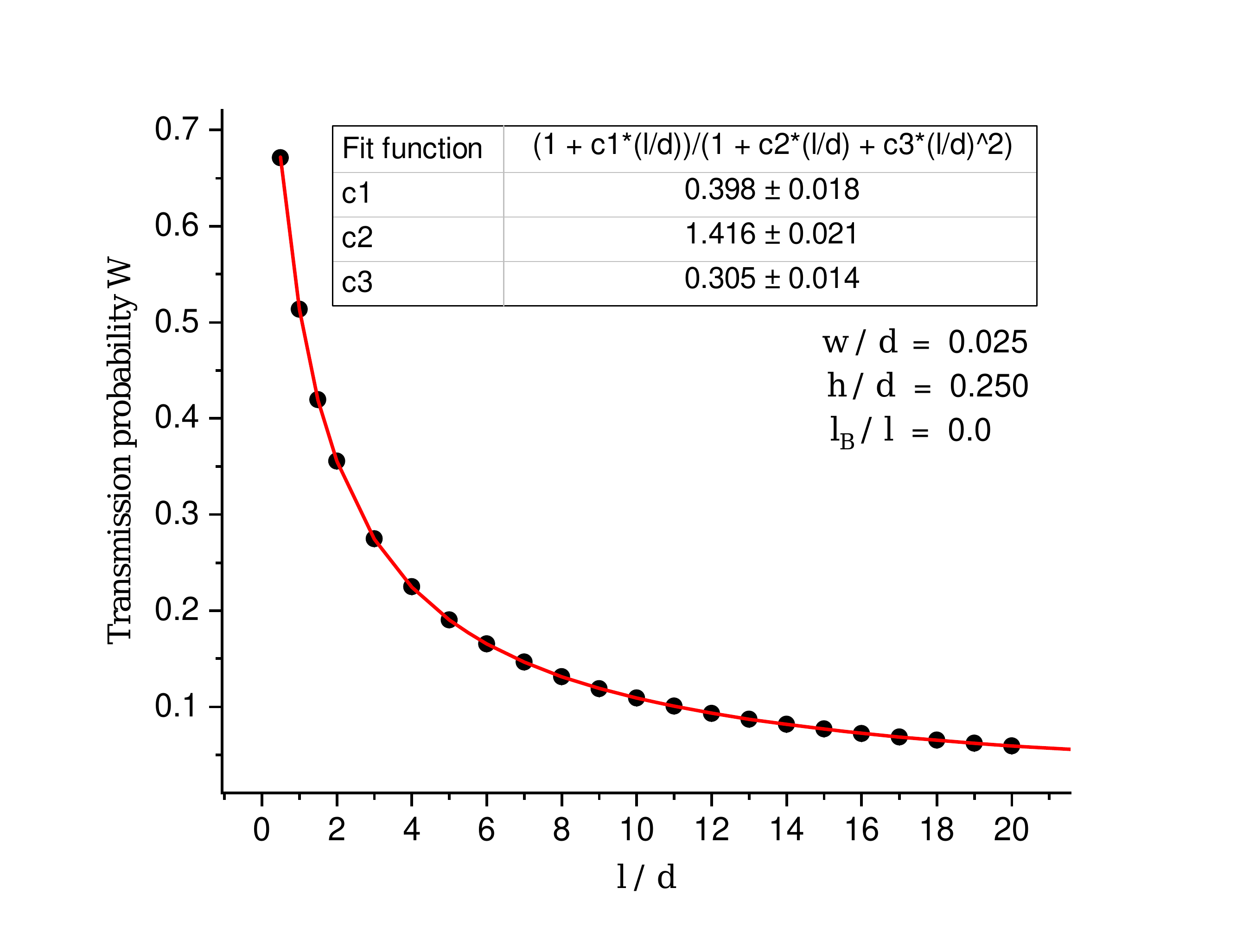} 
\includegraphics[trim=27mm 7mm 33mm 20mm,clip,width=0.48\textwidth]{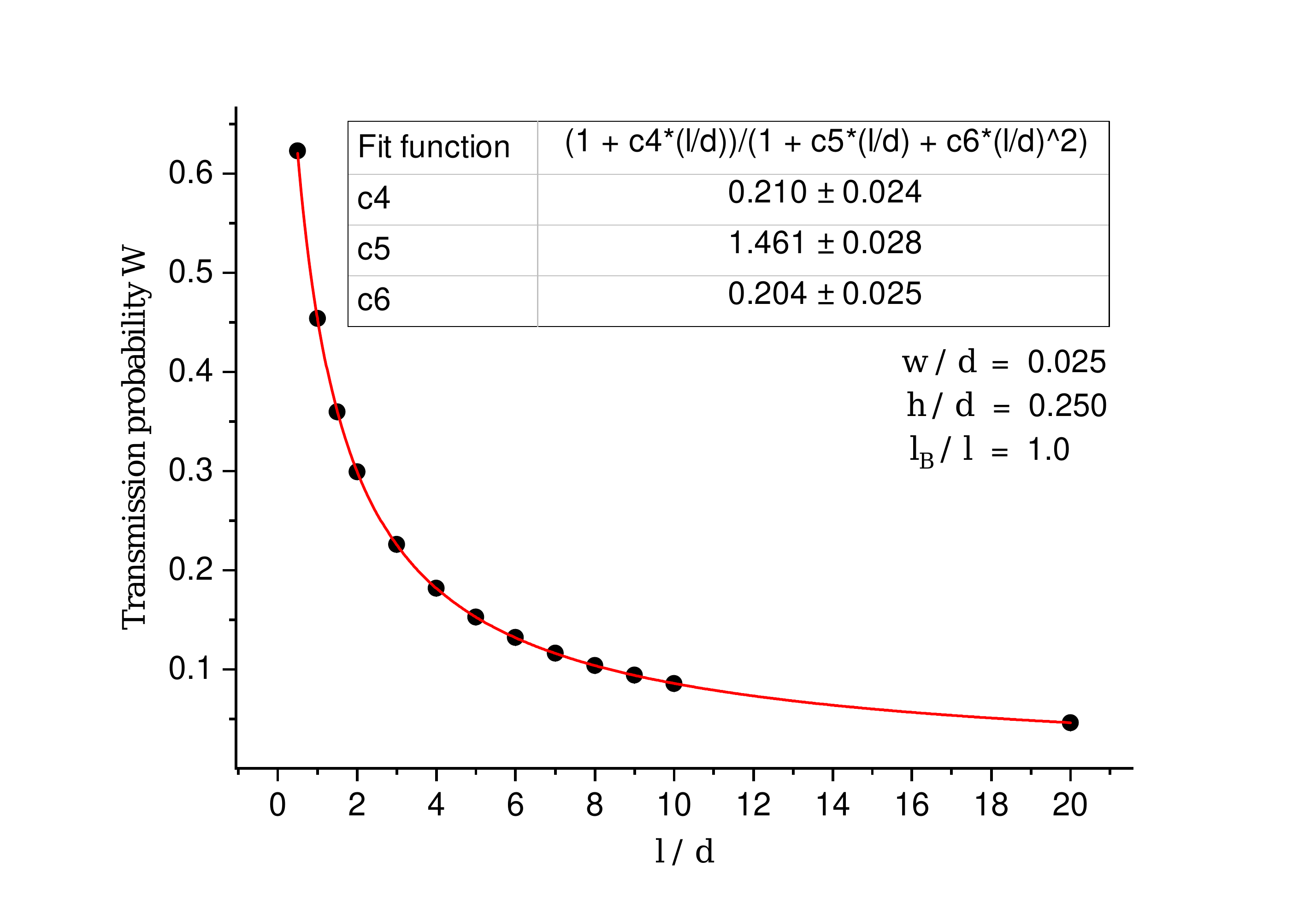} 
\caption{(Top) Fit of Eq.~\ref{eq:ourBellowFormula} on the TPMC simulations of a cylindrical 
tube ($l_B/l = 0$). The black circles represent the simulated transmission probabilities 
$W(l/d)$, plotted over tube length $l/d$. The solid line represents the result of the fit.
(Bottom) Fit of Eq.~\ref{eq:ourBellowFormula} on the TPMC simulations of a pure bellow 
($l_B/l = 1$). Like before, the black circles and the solid line represent the simulated 
transmission probabilities $W(l/d)$ and the result of the fit, respectively. In both cases the 
dependence of the transmission probability on the the length of the tube (bellow) is well described 
by Eq.~\ref{eq:ourBellowFormula}.} \label{fig:resultsStraightTube} 
\end{figure}

\section{Results of the simulations \label{sec:simulationResults}}

In this section, we report the results of the simulations of the different geometries introduced in the previous section. Additionally, a procedure to remove edge-welded bellows from vacuum simulations is proposed, in order to save simulation time without introducing large errors due to the missing bellows.

\subsection{Dependence of the transmission probability on the bellow angle \label{sec:dependenceElementGeometry}}

The first study investigates the dependence of the transmission probability on the geometry of the single bellow elements. The simulated transmission probabilities $W$, together with their statistical error bars, are plotted in Fig.~\ref{fig:resultsAngleDependence} as a function of the bellow element angle $\alpha _B$, defined in Eq.~(\ref{eq:bellowAngle}), for two distinct, fixed values of $l/d$. While the absolute value of $W$ differs for the two different lengths of the tube, the functional dependence of $W$ on $\alpha _B$ is approximately the same for both lengths. A qualitative analysis of the plot allows us to formulate three statements:

\begin{enumerate}
\item The limiting case $\alpha _B = 0$ represents a cylindrical tube. In the vicinity of this value, the transmission probability is approximately independent of $\alpha _B$ within a few percent. Additionally, for large values of $\alpha _B$, the transmission probability is approximately independent of $\alpha _B$ as well.
\item For all values $\alpha _B > 0^\circ$, the transmission probability is lower than the one for $\alpha _B =0$, showing that in general, a bellow has a significantly lower transmission probability than a cylindrical tube of the same length.
\item The transmission probability reaches a minimum for angles in the vicinity of $\alpha _B \approx 55^\circ$.
\end{enumerate}

Qualitatively, one can understand these observations by physical reasoning. Illustrated in 
Fig.~\ref{fig:dependenceOnAngle} is a single bellow element with trajectories of gas particles for (a) a very small bellow angle, (b) a bellow angle of around $\alpha _B \approx 55^\circ$ and (c) a very large bellow angle. In the illustration, the inlet of the bellow is on the left. For small angles, the bellow approximately resembles a cylindrical tube. Hence, the scattering of the gas particles is effectively the same as in the case of a tube. There is no increased backward beaming, since the bellow element acts as a nearly diffuse reflector. This reflective behavior is depicted in gray by the reflection distribution, in this case approximately given by a cosine distribution. For very large angles, the gas particles scatter frequently inside the bellow elements. The net effect of these numerous scattering events is that the gas particles leave the bellow element with an increased beaming towards the opposite wall of the bellow, as indicated by the gray reflection distribution. Due to this, the transmission probability becomes approximately independent of $\alpha _B$ for large angles, since the beaming towards the opposite side of the bellow does not influence $W$ much. The large difference of transmission probabilities between large $\alpha _B$ and $\alpha _B = 0$ can be explained by the reduction of particle trajectories emitted towards the outlet of the bellow, which would otherwise add to the strong forward beaming effect of long cylindrical tubes. Finally, for $\alpha _B \approx 55^\circ$, the backwards beaming becomes maximal for the majority of bellow elements, leading to a minimal transmission probability for this angle, as indicated by the gray distribution function which is directed towards the entrance of the tube. Since the exact position of this minimum critically depends on all geometry parameters of the bellow, we restrict ourselves to this qualitative analysis. 

\subsection{New model for the transmission probabilites of bellows \label{sec:NewModelBellows}}

For typical bellows, the geometry parameters are chosen such that $w/d \ll 1$ and 
$\alpha _B > 80^\circ$. As the qualitative analysis has shown, the transmission probability is 
nearly independent of $\alpha _B$ in such cases, and consequently, the dependency of $W$ on $w$ 
and $h$  for edge-welded bellows is not taken into account in the quantitative model described in 
this section. With fixed values of $w/d = 0.025$ and $h/d = 0.25$, the transmission probability 
$W$ can be fully parametrized as a function of $l/d$ and $l_B/l$. 

\begin{figure}[t]
\centering
\includegraphics[trim=18mm 11mm 35mm 17mm,clip,width=0.48\textwidth]{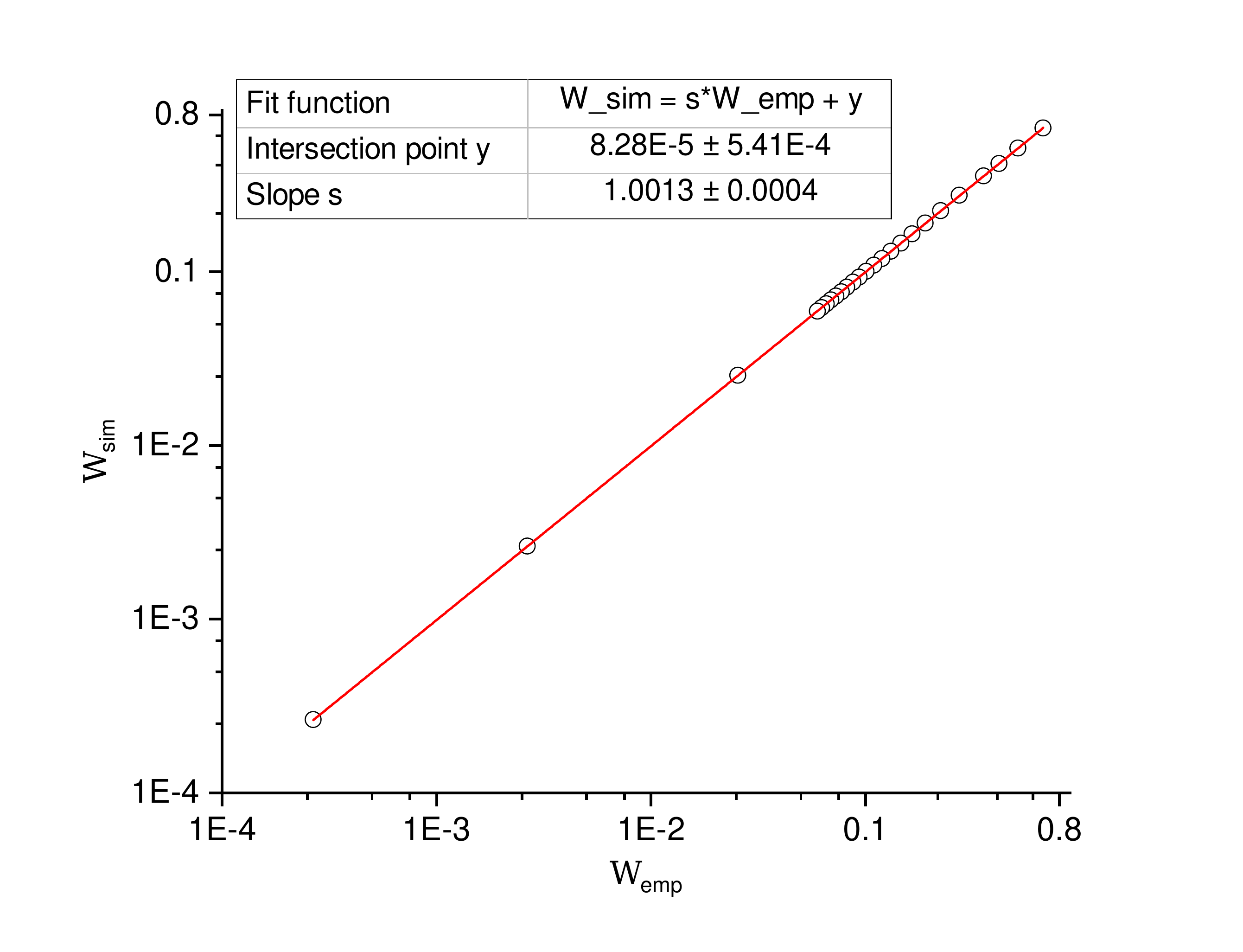} 
\caption{Comparison of the TPMC simulation results with the empirical formula in 
Eq.~(\ref{eq:joustenFormula}). The dots represent the simulated transmission probabilities 
$W_\textrm{sim}$ with $l_B/l = 0.0$ over the transmission probabilities $W_\textrm{emp}$, 
calculated with Eq.~(\ref{eq:joustenFormula}) for various $l/d \in \left[ 0.5 ,\, 5000.0 \right]$. 
The straight line fitted to the points has a slope $1.0013 \pm 0.0004$, demonstrating the excellent 
agreement between Eq.~(\ref{eq:joustenFormula}) and the TPMC simulations for cylindrical tubes.} 
\label{fig:goodnessOfSimulation} 
\end{figure}

For $l_B/l=0.0$ the transmission probability function $W(l/d, l_B/l)$ for an edge-welded bellow 
should turn into Eq.~(\ref{eq:ourBellowFormula}) for a cylindrical tube, which we call $W_0(l/d)$. 
The parameters $c_{1\dots 3}$ in the formula are equivalent to the empirical result from 
Eq.~(\ref{eq:joustenFormula}). For the case of a pure bellow ($l_B / l = 1$) the same ansatz is 
used, called $W_1(l/d)$, but with different parameters $c_{4\dots 6}$. The independent fits of the 
simulation results with Eq.~(\ref{eq:ourBellowFormula}) are shown in 
Fig.~\ref{fig:resultsStraightTube} for both cases. 

In order to test the consistency of the simulated data with the empirical formula in 
\cite{lit:jousten} for a cylindrical tube, we plotted the simulation results for the case 
$l_B/l = 0.0$ over the results of Eq.~(\ref{eq:joustenFormula}). The resulting plot is shown 
in Fig.~\ref{fig:goodnessOfSimulation}. The straight line fitted to the points has a slope 
$1.0013 \pm 0.0004$. This demonstrates the excellent agreement between the widely used 
approximation of Eq.~(\ref{eq:joustenFormula}) and our TPMC simulations for cylindrical tubes.
 
\begin{figure}[t]
\centering
\includegraphics[trim=24mm 9mm 36mm 23mm,clip,width=0.48\textwidth]{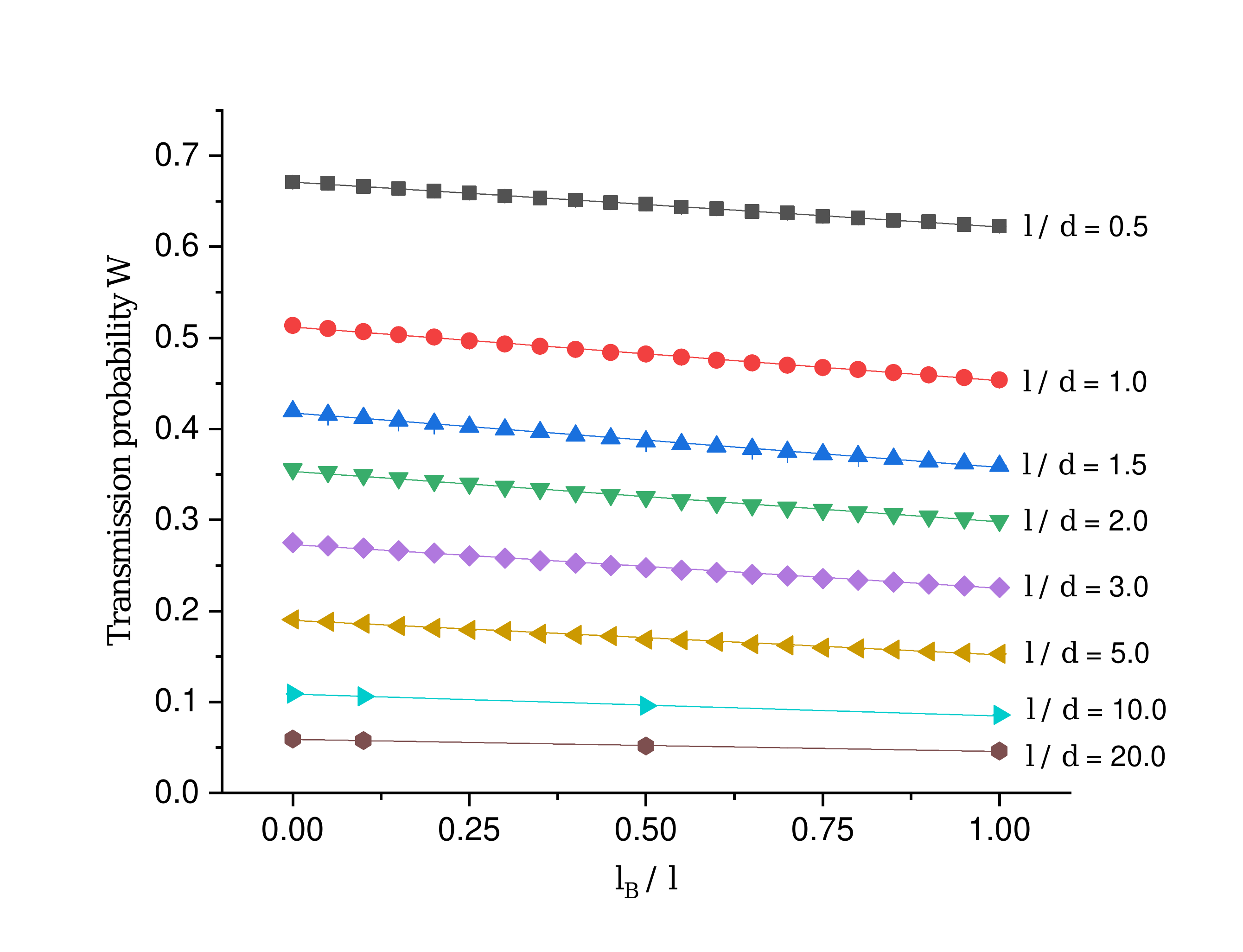} 
\caption{Selection of simulation results for different bellow parameters. Each dot represents the 
simulated transmission probability $W$ for a fixed set of geometry parameters $l/d$ and $l_B/l$. 
The statistical errors of the simulation results are of $\mathcal{O} (10^{-4})$. Therefore, the 
error bars are not visible for most of the points. For fixed values of $l/d$ the linear dependence 
of $W$ on $l_B/l$ is indicated by straight lines. The lines represent the fit results of 
Eq.~(\ref{eq:ourFullBellowFormula}), with the parameters $c_i$ given in 
Tab.~\ref{tab:regressionParameterResults}.} \label{fig:resultsWForDifferentLengths} 
\end{figure}

The dots in Fig. \ref{fig:resultsWForDifferentLengths} represent the simulated 
transmission probabilities $W(l/d, l_B/l)$ for a variaty of geometry parameters 
$l/d$ and $l_B/l$. For constant values of $l/d$ the simulation results show a linear 
drop between $W_0(l/d)$ and $W_1(l/d)$. Thus, for arbitrary values of 
$l_B/l \in \left[0 ,\,1\right]$, the transmission probability can be interpolated by a 
linear equation:

\begin{eqnarray}
 W(l/d, l_B/l) &=& W_0(l/d) + \frac{l_b}{l}\cdot \left( W_1(l/d) - W_0(l/d)\right) \nonumber \\
               &=& \left( 1-\frac{l_B}{l}\right)\cdot\frac{1 + c_1\cdot\frac{l}{d}}{1 + c_2\cdot 
                   \frac{l}{d} + c_3\cdot\left( \frac{l}{d} \right) ^2}
                   \label{eq:ourFullBellowFormula}\\
			   &\hspace*{0.4cm}& + \left( \frac{l_B}{l}\right)\cdot\frac{1 + 
              c_4\cdot\frac{l}{d}}{1 + c_5\cdot\frac{l}{d} + 
              c_6\cdot \left( \frac{l}{d} \right) ^2} \nonumber
\end{eqnarray}

\noindent The full set of parameters $c_{1\dots 6}$ has been determined by fitting 
Eq.~(\ref{eq:ourFullBellowFormula}) to all of the simulated data. The results for the six 
parameters are presented in Tab.~\ref{tab:regressionParameterResults}, together with their 
statistical errors. The straight lines in Fig.~\ref{fig:resultsWForDifferentLengths} have been 
calculated with Eq.~(\ref{eq:ourFullBellowFormula}). They connect the simulated points which share 
the same value of $l/d$. The individual simulations are in very good agreement with this empirical 
model. Due to the high number of simulated particles, the error of each result is of 
$\mathcal{O} (10^{-4})$. For most of the points the error bars are too small to be visible. 
For values of $l/d > 10$ the number of different $l_B / l$ simulations was restricted to 
fewer values, due to the rapidly increasing computation time. For the same reason the high 
statistics of adsorbed particles of $\mathcal{O} (10^6)$ could not be maintained for larger 
values of $l_B / l$. It had to be reduced to $\mathcal{O} (10^5)$ adsorptions. The simulation of 
the single point at $W(l/d = 5000, l_B/l = 0)$ was stopped due to the long simulation time, after reaching a statistical uncertainty of better than 1\%.

\begin{table}[t]
    \caption{Parameters $c_i$ and their statistical errors of Eq.~(\ref{eq:ourFullBellowFormula}), 
    fitted to the simulated transmission probabilities $W(l/d, l_B/l)$.}
    \centering
    \label{tab:regressionParameterResults}
	\begin{tabular}{ c c c }
  \hline			
  Parameter & Fit value & Standard error \\
  \hline  
  $c_1$ &  0.2915 & 0.0020 \\
  $c_2$ &  1.3018 & 0.0026 \\
  $c_3$ &  0.2225 & 0.0015 \\
  $c_4$ &  0.2707 & 0.0069 \\
  $c_5$ &  1.5267 & 0.0077 \\
  $c_6$ &  0.2777 & 0.0077 \\
  \hline  
	\end{tabular}
\end{table}

For $l \ll d$ the ratio $W_1(l \ll d)/W_0(l \ll d)$ of the transmission probabilities of a 
pure bellow and a cylindrical tube, parametrized by Eq.~(\ref{eq:ourFullBellowFormula}), 
converges against a constant value

\begin{eqnarray}
 \frac{W_1(l \ll d)}{W_0(l \ll d)} = \frac{c_3\cdot c_4}{c_1\cdot c_6} = 0.74 \pm 0.03
 \label{eq:convergence}
\end{eqnarray}

\noindent For cylindrical tubes the transmission probability $W_0$ converges towards $(1.322 \pm 0.013)\cdot d/l$ for $l\gg d$, which agrees well with the literature value of Eq.~(\ref{eq:clausinFormula2}). The transmission probability $W_1$ for pure bellows converges towards $(0.98 \pm 0.04)\cdot d/l$ for $l\gg d$. 
 
\subsection{Comparison of simulation time for bellows and tubes \label{sec:ComputationTime}}

\begin{figure}[t] 
\centering
\includegraphics[trim=24mm 10mm 38mm 20mm,clip,width=0.48\textwidth]{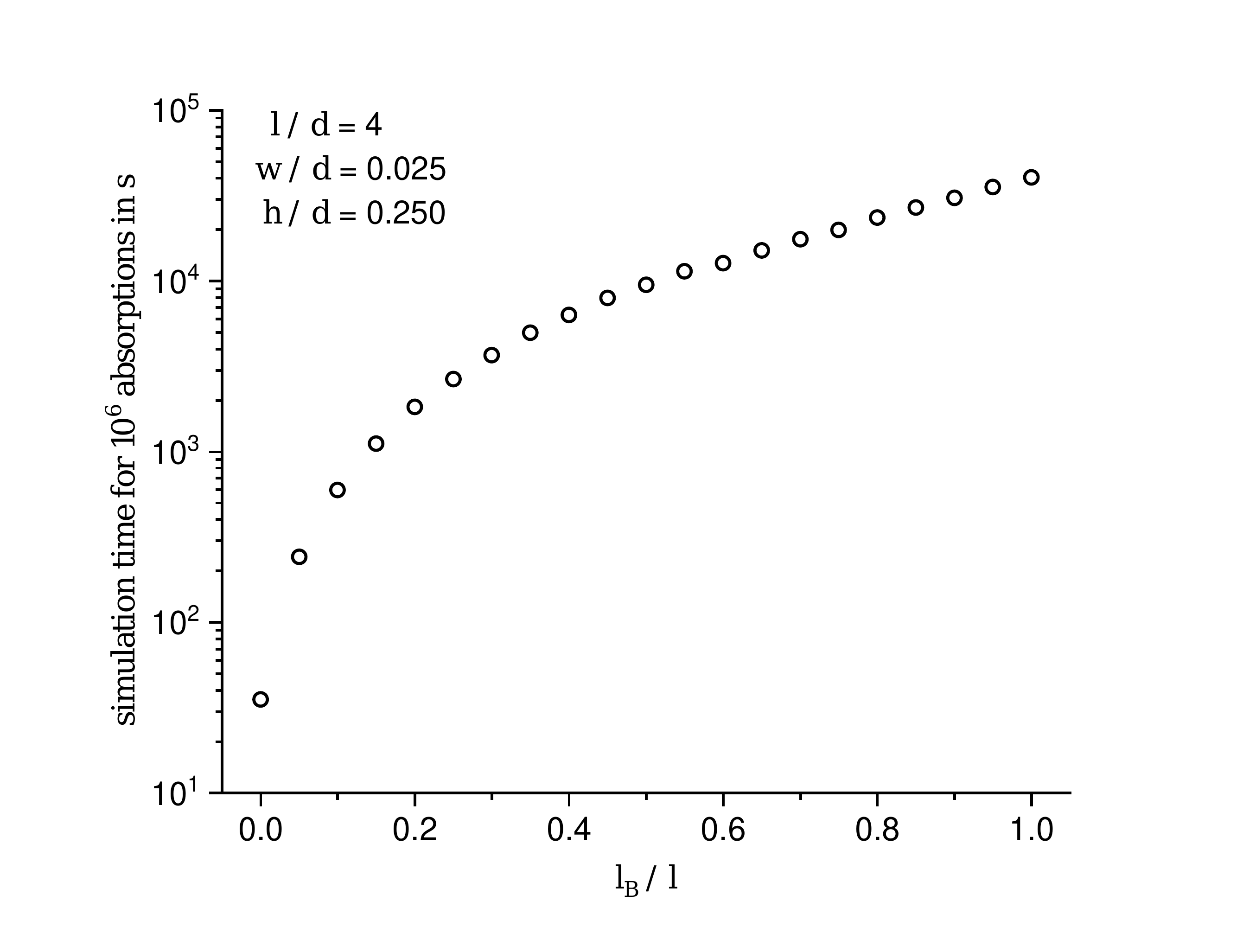} 
\caption{Effect of the modeling of the geometry of a bellow on the simulation time. Plotted is the 
time needed for the simulation of $N_A=10^6$ adsorptions on the outlet surface of the bellow in 
seconds for the entire range of $l_B/l \in \left[ 0.0 ,\, 1.0\right]$ for fixed values of 
$l/d = 4$, $w/d = 0.025$ and $h/d = 0.25$. The simulation time increases by more than three 
orders of magnitude compared to a cylindrical tube ($l_B/l = 0$).} \label{fig:SimulationTime} 
\end{figure}

In order to investigate the effect of accurately modeled edge-welded bellows in a TPMC simulation
 with respect to the simulation time, the geometry parameter set $l/d = 4$, $w/d = 0.025$ and 
$h/d = 0.25$ is used as a benchmark. For values $l_B/l \in \left[ 0.0 ,\, 1.0\right]$ with 
steps of $0.05$, each bellow model was simulated until the number of adsorptions reached 
$N_A = 10^6$ on the outlet surface of the bellow. The required simulation time is shown in 
Fig.~\ref{fig:SimulationTime}. As can be seen in the logarithmic plot, the simulation time  
considerably increases with the length of the bellow segments, represented by the fraction 
$l_B / l$. For each additional bellow element, the number of facets in the simulation 
increases, as well as the time a particle spends trapped inside the narrow bellow elements. This 
leads to a non-linear increase of the simulation time. For values of $l/d > 10$ with 
$l_B /l \in \left[ 0.5 ,\, 1.0\right]$ a simulation was not practicable anymore with the same 
accuracy, since the time required to reach $10^6$ adsorptions was too long. This analysis shows 
that bellows have a significant effect on the simulation time, which can rise by several orders of 
magnitudes.

\subsection{Replacement of a bellow with an adapted cylindrical tube \label{sec:ReplacementBellow}}
Due to the substantial increase in CPU time when modeling an edge-welded bellow in a TPMC 
simulation, we suggest a method to optimize vacuum simulations without a reduction of the 
statistical significance of the result. The transmission probability $W$ of a bellow in a vacuum 
system can be estimated by Eq.~(\ref{eq:ourFullBellowFormula}). Since the simulation of a 
cylindrical tube with the same transmission probability $W$ as the bellow is much faster, one can 
replace the bellow by a slightly longer cylindrical tube with an effective length $l'$. The 
extended length would compensate the larger conductance of the tube with regard to the bellow. 
With the \textit{ansatz} $W = W(l/d, l_B/l) = W(l'/d, 0)$, one finds for the effective length $l'$   

\begin{equation}
	\frac{l'}{d} = \frac{c_1 - c_2\, W + \sqrt{(c_1 - c_2\, W)^2 + 4\,c_3\,W\,(1-W)}}{2\,c_3\,W} ~.
\label{eq:invertedBellowFormula}
\end{equation}

\noindent  by inverting Eq.~(\ref{eq:ourFullBellowFormula}) for $l_B / l=0$.
Thus, for reducing the simulation time for vacuum setups containing edge-welded bellows, 
we propose the following approach:

\begin{enumerate}
\item For each bellow in the vacuum setup, calculate its approximate transmission probability $W$ 
with Eq. (\ref{eq:ourFullBellowFormula}), using the parameters in 
Tab.~\ref{tab:regressionParameterResults}.
\item Calculate the effective length $l'$ of a cylindrical tube with the same transmission probability $W$ as the bellow, using Eq.~(\ref{eq:invertedBellowFormula}).
\item Replace the bellow in the TPMC model with a cylindrical tube of length $l'$.
\end{enumerate}

\noindent The proposed substitution changes the angular distribution of the trajectories of the particles leaving the bellow. In long tubes the well known beaming effect is more pronounced than in bellows. Therefore, this method is most accurate if the bellow connects to vacuum chambers with larger dimensions than the diameter of the bellow, where the beaming effect plays only a minor role in the subsequent distribution of the simulated particles. 

The transmission time through a tube is faster than the transmission time through a bellow, where the particles spend more time in the crevices of the bellow segments. Most TPMC simulations investigate the state of a vacuum system that has already reached equilibrium. However, if the dynamic development of the pressure over time is an issue, the proposed method should be used with a grain of salt. 

Apart from these two special cases the proposed substitution of a bellow by a cylindrical tube with an appropriate length $l'$ can lead to a considerable reduction of the simulation time.   

\section{Conclusions \label{sec:conclusions}}

We have simulated the transmission probability $W$ of edge-welded bellows for a wide range of design parameters with the TPMC code MolFlow+. An empirical model was developed to describe the results in a concise way.

The analysis shows that for typical edge-welded bellows, the geometrical parameters of the individual bellow elements, or more specifically, the bellow angle $\alpha _B$, has only a minor influence on the transmission probability. In most applications its influence can be neglected, leading to a quantitative description, which only depends on the ratio of the length and diameter of the bellow. 

The quantitative model of $W$, presented in this work, examined the more general case, where the bellow features cylindrical tubes at both ends. It is parameterized with the total length over diameter $l/d$ and the relative length $l_B/l$ of the bellow. For a fixed total length $W$ drops linearly with the increasing share $l_B/l$ of the bellow section. For fixed $l_B/l$, the dependence of $W$ on $l/d$ follows the functional form of the commonly used empirical formula (Eq.~\ref{eq:joustenFormula}) of the transmission probability for a cylindrical tube. The results from a large number of simulations with different bellow and tube lengths have been used to determine the six parameters of the model. 

For very long bellows the model converges towards a constant ratio $W_1/W_0 = 0.74 \pm 0.03$ between the transmission probabilities of a bellow and a cylindrical tube. With the commonly used approximation $W_0 = 4\,d/3\,l$ for very long tubes, the transmission probability of long bellows can be estimated as $W_0 = d/l$ within the errors of the simulations. 

In an attempt to reduce the prohibitively long TPMC simulation times for long bellows, a method is proposed, where the empirical model is used to calculate the appropriate length of a cylindrical tube, which replaces the bellow in the TPMC model. This method can reduce the simulation time by several orders of magnitude, while preserving the statistical significance of the simulation.

\section*{Acknowledgment}
We want to thank Roberto Kersevan and Marton Ady from CERN for making the source code of MolFlow+ 2.5 available to us. Additionally, we want to thank the KATRIN collaboration for supporting this work.

\end{document}